# Tunneling Electroresistance
# in Ferroelectric Tunnel Junctions with a Composite Barrier

M. Ye. Zhuravlev[1,2], Y. Wang[1], S. Maekawa[3,4], and E. Y. Tsymbal[1*]

[1] *Department of Physics and Astronomy, Nebraska Center for Materials and Nanoscience,
University of Nebraska, Lincoln, Nebraska 68588-0111, USA*

[2] *Kurnakov Institute for General and Inorganic Chemistry, Russian Academy of Sciences, 119991 Moscow, Russia*

[3] *Institute for Materials Research, Tohoku University, Sendai 980-8577, Japan*

[4] *CREST, Japan Science and Technology Agency, Sanbancho, Tokyo 102-0075, Japan*

Tunneling electroresistance (TER) effect is the change in the electrical resistance of a ferroelectric tunnel junction (FTJ) associated with polarization reversal in the ferroelectric barrier layer. Here we predict that a FTJ with a composite barrier that combines a functional ferroelectric film and a thin layer of a non-polar dielectric can exhibit a significantly enhanced TER. Due to the change in the electrostatic potential with polarization reversal the non-polar dielectric barrier acts as a switch that changes its barrier height from a low to high value. The predicted values of TER are giant and indicate that the resistance of the FTJ can be changed by many orders in magnitude at the coercive electric field of ferroelectric.

Thin-film ferroelectric materials have recently attracted significant interest due to their technological application in electronic devices such as ferroelectric memories.[1] One of the critical characteristics affecting the performance of memory devices based on ferroelectric capacitors is leakage currents. Therefore, conduction through thin ferroelectric films has been active area of research over the years (see, e.g., ref. 2). For films of hundred nanometer thickness it was found that transport mechanisms are similar to those known for wide band gap semiconductors. In particular, Schottky thermionic emission,[3] Poole-Frenkel conduction,[4] and Fowler-Nordheim tunneling[5] were considered as possible sources of leakage currents in ferroelectric capacitors.[6]

This transport behavior changes dramatically when a film thickness approaches a nanometer scale, causing direct tunneling to be the dominant mechanism of conduction.[7] Recent experimental[8] and theoretical[9] studies of perovskite ferroelectric oxides have demonstrated that ferroelectricity persists down to a nanometer scale, which makes it possible to use ferroelectrics as tunnel barriers in ferroelectric tunnel junctions (FTJs).[7] Contrary to ferroelectric capacitors where leakage currents are detrimental to the device performance, the conductance of a FTJ is the functional characteristic of the device. The key property is tunneling electroresistance (TER) that is the change in resistance of a FTJ with reversal of ferroelectric polarization. Based on simple models it was predicted that TER in FTJs can be sizable due to the change in the tunneling potential barrier dependent of polarization orientation.[10,11] These results were elaborated using first-principles calculations of transport properties of FTJs.[12,13] Indications of the TER effect have been seen in experiments on Pt/Pb(Zr$_{0.52}$Ti$_{0.48}$)O$_3$/SrRuO$_3$ junctions and more recently in FTJs utilizing a multiferroic La$_{0.1}$Bi$_{0.9}$MnO$_3$ barrier.[14] Very recently a giant TER was unambiguously demonstrated using local probe measurements on nm-thick BaTiO$_3$ films.[15,16]

There are several mechanisms responsible for the TER effect in FTJs.[7] One of them involves different screening lengths in metallic electrodes[10] and two others originate from the transmission across the interface affected by polarization orientation[12] and the polarization dependent decay constant in a ferroelectric barrier.[13] All the mechanisms require asymmetry in a FTJ, which may be intrinsic (e.g., due to non-equivalent interfaces) or intentionally introduced in the system (e.g., by using different electrodes).

In this letter we propose an efficient way to enhance the TER considerably by using a layered composite barrier combining a functional ferroelectric film and a thin film of a non-polar dielectric material. Due to the change in the electrostatic potential induced by polarization reversal the non-polar film adjacent to one of the interfaces acts as a switch changing its barrier height from a low to high value resulting in a dramatic change in the transmission across the FTJ. The predicted values of TER are giant, indicating that the resistance ratio between the two polarization-orientation states in such FTJs may reach hundred thousands and even higher. The proposed geometry of a FTJ does not require different electrodes and may be practical for device application.

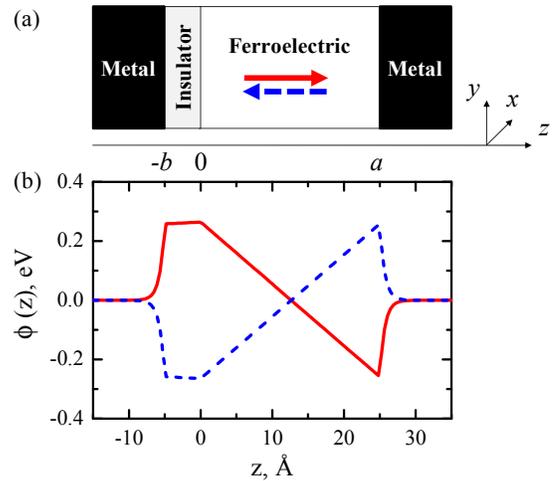

**Fig. 1:** Geometry (a) and the electrostatic potential profile for the two opposite polarization orientations (b) of a ferroelectric tunnel junction: $a = 25$Å, $b = 5$Å, $\varepsilon_d = 300$, $\varepsilon_f = 90$, $\delta = 1$Å, and $P = 20$ μC/cm$^2$.

Following Ref. 10, we employ a free-electron model to describe the electrostatic potential and electron tunneling in a FTJ. The overall tunneling potential profile seen by transport electrons is assumed to be a superposition of the step-wise potential originating from the variation of the conduction band minimum across the FTJ and the electrostatic potential resulting from the spontaneous uniform polarization of the ferroelectric layer, the induced polarization in the non-polar dielectric layer and the screening charge in the electrodes.

To calculate the electrostatic potential we employ the Thomas-Fermi model of screening and impose short-circuit boundary conditions. In this case the shape of the electrostatic potential in the electrodes is given by

$$\phi(z) = \begin{cases} \varepsilon_0^{-1}\sigma_s\delta\exp((z+b)/\delta), & z < -b \\ -\varepsilon_0^{-1}\sigma_s\delta\exp(-(z-a)/\delta), & z > a \end{cases} \quad (1)$$

Here $\sigma_s$ is the screening charge, $\delta$ is Thomas-Fermi screening length in the electrodes, which are assumed to be identical, and $a$ and $b$ are thicknesses of the ferroelectric and dielectric layers (see Fig. 1a). The potential at the interfaces and the screening charge can be found from the boundary conditions which lead to:

$$\sigma_s = \frac{a\varepsilon_d\varepsilon_0^{-1}P}{a\varepsilon_d + b\varepsilon_f + 2\varepsilon_f\varepsilon_d\delta}$$

$$\phi(0) = \frac{(\delta\varepsilon_d + b)a\varepsilon_0^{-1}P}{a\varepsilon_d + b\varepsilon_f + 2\varepsilon_f\varepsilon_d\delta} \quad (2)$$

$$\phi(-b) = -\phi(a) = \frac{a\delta\varepsilon_d\varepsilon_0^{-1}P}{a\varepsilon_d + b\varepsilon_f + 2\varepsilon_f\varepsilon_d\delta}$$

Here $P$ is the ferroelectric polarization and $\varepsilon_f$ and $\varepsilon_d$ are the dielectric constants of the ferroelectric and dielectric layers respectively.

Fig. 1b shows the calculated electrostatic potential profile across the FTJ, assuming that $\delta = 1$Å, $P = 20$ μC/cm$^2$, $\varepsilon_f = 90$, and $\varepsilon_d = 300$. These values approximately describe the dielectric properties of ferroelectric BaTiO$_3$[17] and non-polar dielectric SrTiO$_3$. As seen from Fig. 1b, the electrostatic potential changes by about 0.5V across the ferroelectric layer reflecting the presence of the depolarizing field.[18] As follows from Eqs. (2), the depolarizing field is non-vanishing even when screening in the electrodes is perfect, i.e. $\delta = 0$, as long as the non-polar dielectric layer has finite thickness. When polarization switches the depolarizing field reverses its direction (Fig. 1) and alters the electrostatic potential in the dielectric layer by about 0.5eV. This change is significant to produce a strong TER effect.

The tunneling conductance of FTJs per unit area is calculated using the Landauer formula,

$$G = \frac{2e^2}{h}\int \frac{d^2\mathbf{k}_\parallel}{(2\pi)^2}T(E_F,\mathbf{k}_\parallel), \quad (3)$$

where $T(E_F,\mathbf{k}_\parallel)$ is the transmission coefficient at the Fermi energy $E_F$ for a given value of the transverse wave vector $\mathbf{k}_\parallel$. The transmission coefficient is obtained from the solution of the Schrödinger equation for an electron moving in the potential $V(z)$ by imposing a boundary condition of the incoming plane wave normalized to a unit flux density and by calculating the amplitude of the transmitted plane wave. The solution is obtained numerically for the potential $V(z)$ which is the superposition of the electrostatic potential $\phi(z)$ and the step-wise potential originating from the variation of the conduction band minimum across the FTJ. For a given Fermi energy $E_F$ in the metal electrodes, the latter determines the barrier heights $U_d$ and $U_f$ for the non-polar dielectric and ferroelectric layers respectively. We assume that electrons have a free electron mass, the Fermi energy is $E_F = 3$eV, and the ferroelectric barrier height is $U_f = 0.6$eV. We define the TER ratio as follows: TER=$(G_L - G_R)/G_R$, where $G_L$ ($G_R$) is the conductance of a FTJ for polarization pointing left (right).

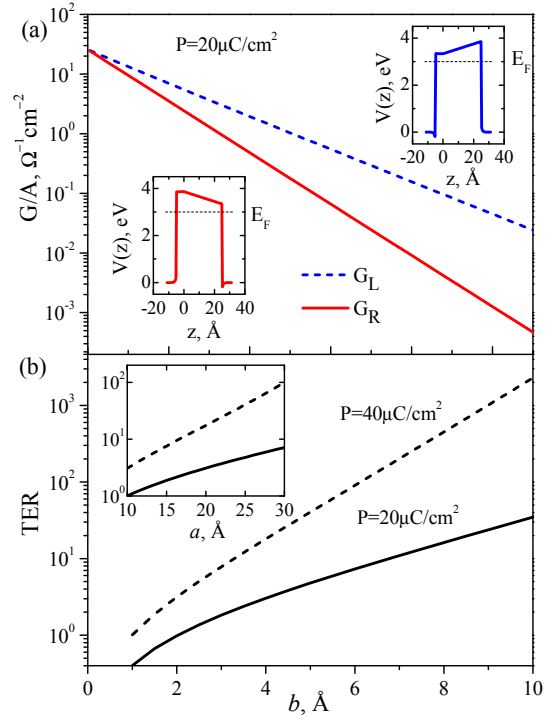

**Fig. 2:** (a) Conductance of a FTJ for two opposite polarization orientations: left (solid line) and right (dashed line), as a function of dielectric layer thickness. The insets show the corresponding tunneling barrier profiles. (b) TER as the function of dielectric layer thickness for two polarizations $P = 20$ μC/cm$^2$ (solid line) and $P = 40$ μC/cm$^2$ (dashed line). The inset shows TER as the function of the ferroelectric film thickness. $U_d$=0.6 eV, $\varepsilon_d$= 300.

Calculations of the conductance and TER are performed for two systems which are dissimilar by a different non-polar dielectric layer. In the first case we consider a dielectric that is described by the barrier height $U_d = 0.6$ eV and the dielectric constant $\varepsilon_d = 300$, which is an approximation for a SrTiO$_3$ barrier. In the second case we consider a dielectric that is described by $U_d = 2.5$ eV and $\varepsilon_d = 10$, which is an approximation for a MgO barrier.

Fig. 2a shows the conductance per unit area for the two opposite directions of the ferroelectric polarization as a function of the dielectric layer thickness $b$, in case of SrTiO$_3$/BaTiO$_3$ composite barrier. Due to identical electrodes, when $b \to 0$ the difference between $G_L$ and $G_R$ vanishes and TER=0. With increasing $b$ both $G_L$ and $G_R$ decrease exponentially. However, the slope of the $G(d)$ dependence is different for the two opposite polarization orientations, reflecting a different barrier height (see the insets in Fig. 3a) and consequently a different decay constant. As a result the TER ratio increases exponentially with the dielectric layer thickness (Fig. 2b). The exponential enhancement of TER is also predicted as a function of the ferroelectric layer thickness $a$ (see the inset in Fig. 2b). This is the consequence of the electrostatic potential drop across the ferroelectric layer which increases linearly with $a$ and leads to the enhanced asymmetry of the tunneling barrier created by the dielectric layer for the two polarization orientations. The latter effect is also responsible for the TER enhancement with polarization of the ferroelectric layer (see Fig. 2 b).

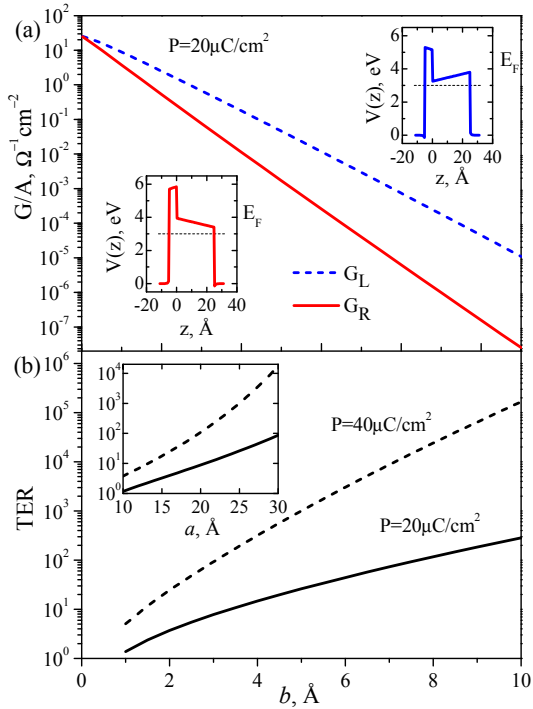

**Fig.3:** Same as in Fig. 2 for $U_d$ =2.5 eV and $\varepsilon_d$ = 10.

A qualitatively similar behavior of the conductance and TER is predicted for MgO/BaTiO$_3$ composite barrier (see Fig. 3). However, due to a larger amplitude of the tunneling barrier in the dielectric ($U_d$ = 2.5 eV), the conductance changes more steep as a function of the dielectric barrier thickness (Fig. 3a), leading to even larger values of TER (Fig. 3b). As is evident from Fig. 3b, for reasonable values of the dielectric and ferroelectric barrier thickness the TER ratio is really giant, reaching hundred thousands and even larger.

Recent experiments on TER in ultrathin ferroelectric films indicate that very large values of TER may be achieved using local transport measurements on a thin ferroelectric film which is deposited on a metal layer and probed by a conductive AFM tip.[15,16] These results may be the consequence of a native dielectric layer which is created when the ferroelectric film is exposed to air. The results of the present work suggest that depositing a dielectric layer in a controllable way may provide a route to tailor TER.

In conclusion, we have shown that a thin non-polar dielectric layer at the interface between a ferroelectric barrier and a metal electrode in a FTJ may significantly enhance the TER effect. This dielectric layer serves as a switch that changes its barrier height from a low to high value when the polarization of the ferroelectric barrier is reversed, resulting in giant TER values. The proposed method of enhancing TER may be practical for device applications of FTJs.

M.Y.Z. is grateful to Anatoly Vedyayev for valuable discussions. E.Y.T. is thankful to the Institute of Materials Research at Tohoku University for hospitality during his visit in May-June 2009. This work was supported by NSF MRSEC (Grant No. 0820521), the Nanoelectronics Research Initiative of SRC, and the Nebraska Research Initiative.

* e-mail: tsymbal@unl.edu